\documentclass[preprint,showpacs,amsmath,amssymb,showkeys,endfloats]{revtex4}

\usepackage{graphicx}
\usepackage{dcolumn}
\usepackage{bm}
\usepackage{ae,aeguill,aecompl}

\begin{document}

\preprint{APS/123-QED}

\title{Characterization of a CCD array for Bragg spectroscopy}

\author{Paul Indelicato}
 \affiliation{
   Laboratoire Kastler Brossel, Unit{\'e} Mixte de Recherche du CNRS n$^\circ$ 8552,  Universit{\'e} Pierre et Marie Curie,
   Case 74, 4, Place Jussieu, F-75005 Paris, France.
 }
\author{Eric-Olivier Le~Bigot}
 \affiliation{
   Laboratoire Kastler Brossel, Unit{\'e} Mixte de Recherche du CNRS n$^\circ$ 8552,  Universit{\'e} Pierre et Marie Curie,
   Case 74, 4, Place Jussieu, F-75005 Paris, France.
 }
\author{Martino Trassinelli}
\altaffiliation{Corresponding author. Electronic address: \texttt{martino.trassinelli@spectro.jussieu.fr}.}
 \affiliation{
   Laboratoire Kastler Brossel, Unit{\'e} Mixte de Recherche du CNRS n$^\circ$ 8552,  Universit{\'e} Pierre et Marie Curie,
   Case 74, 4, Place Jussieu, F-75005 Paris, France.
 }
\author{Detlev Gotta}
 \affiliation{
   Institut f{\"u}r Kernphysik, Forschungszentrum J{\"u}lich, 
   D-52425 J{\"u}lich, Germany
 }
\author{Maik Hennebach}
 \affiliation{
   Institut f{\"u}r Kernphysik, Forschungszentrum J{\"u}lich, 
   D-52425 J{\"u}lich, Germany
 }
\author{Nick Nelms}
 \affiliation{
   Space Research Center, Department of Physics and Astronomy, University of Leicester,
   University road, Leicester LE1 7RH, United Kingdom.
 }
\author{ Christian David}
 \affiliation{
   Paul Scherrer Institut, CH-5232 Villigen, Switzerland 
 }
\author{ Leopold M. Simons}
 \affiliation{
   Paul Scherrer Institut, CH-5232 Villigen, Switzerland 
 }
%
 

\date{\today}

\begin{abstract}
The average pixel distance as well as the relative orientation of an array of 6 
CCD detectors have been measured with accuracies of about 0.5~nm 
and 50~$\mu$rad, respectively. Such a precision satisfies the needs of modern 
crystal spectroscopy experiments in the field of exotic atoms and highly charged ions.
Two different measurements have been performed by illuminating masks
in front of the detector array by remote sources of radiation. In one case,
an aluminum mask was irradiated with X-rays and in a second attempt,  a 
nanometric quartz wafer was illuminated by a light bulb. Both methods 
gave consistent results with a smaller error for the optical method. 
In addition, the thermal expansion of the CCD detectors was characterized 
between $-105 ^\circ $C and $-40 ^\circ $C. 
\end{abstract}

\maketitle

\section{Introduction}
Charge--coupled devices (CCDs) are ideally suited as detectors for X--ray spectroscopy in the few keV range,  
because of excellent energy resolution and the inherent two-dimensional spatial information.
In particular, they can be used as focal-plane detectors of Bragg crystal spectrometers for studies of 
characteristic X--radiation from exotic atoms with ultimate energy resolution\,\cite{Gotta2004}.  

\begin{figure}
\includegraphics[width=0.4\textwidth]{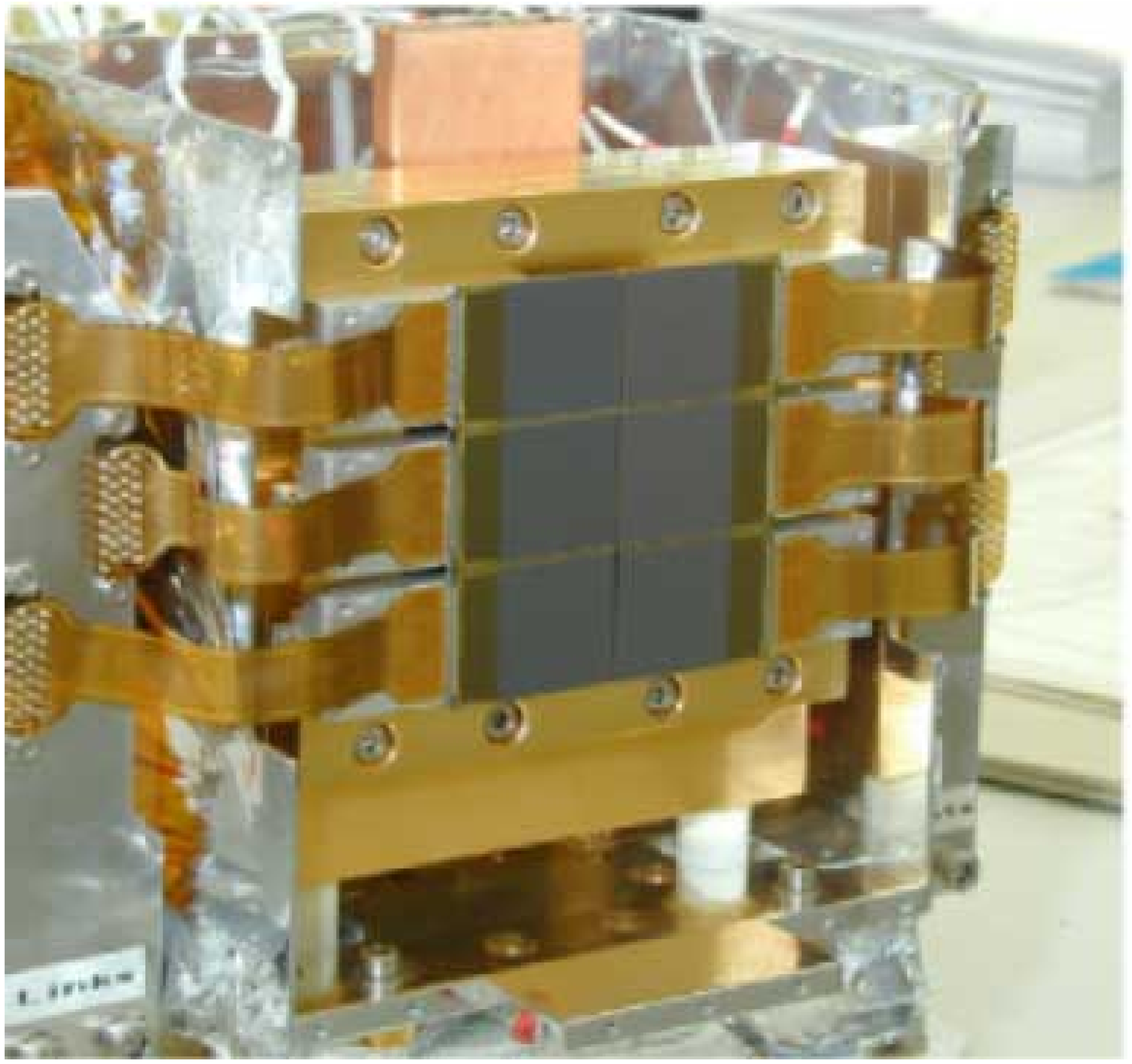}
\caption{Array of 6 CCD devices mounted on the cold head\,\cite{Nelms2002b}.}
\label{array6}
\end{figure}

The detector described in this work was set--up for a bent crystal 
spectrometer used in three ongoing experiments at the Paul Scherrer Institut:
the measurement of the charged pion mass \cite{ProposalPM,Nelms2002a}, 
the determination of the strong--interaction shift and width of the  pionic hydrogen ground state 
\cite{ProposalPH,Anagnostopoulos2003a} and highly charged ion spectroscopy \cite{Trassinelli2005}. The detector is made of an array of 
two vertical columns of 3 CCDs each\,\cite{Nelms2002b} (Fig.\,\ref{array6}).
Each device has $600 \times 600$ square pixels with a nominal dimension of $40~\mu$m at room temperature.
Each pixel is realized by an open-electrode structure. 
For this reason, the dimension characterizing the detector is rather the average distance  
between pixels centers than the size of the individual pixel.

As the CCD is usually operated at $-100 ^\circ $C, the knowledge of 
the inter--pixel distance at the working temperature is essential for crystal spectroscopy, 
because any  angular difference is determined from a measured position difference between
Bragg reflections. Furthermore, for an array like the one described here, the relative orientation 
of the CCDs has to be known at the same level of accuracy as the average pixel distance.

A first attempt to determine the relative positions has been made using a
wire eroded aluminum mask illuminated by sulphur fluorescence X--rays produced by means
of an X--ray tube. The alignment of the mask pattern made it possible to estimate
the relative CCD position with an accuracy of about 0.05\,--\,0.1 pixel and the relative rotation to 
slightly better than 100~$\mu $rad \cite{Hennebach2004}.
In order to obtain in addition a precise value for the average pixel distance 
a new measurement was set--up using a high-precision quartz wafer 
in front of the CCD illuminated with visible light.
Using this method, the relative CCD devices' position was evaluated with an accuracy of about 0.02~pixel.
The temperature dependence of the pixel distance was also determined.

Section~\ref{sec:set-up} is dedicated to the description of the optical measurement set--up.
In section~\ref{sec:pxd}, we describe the measurement of the pixel distance. 
Section~\ref{sec:CCD-orientation} we present the measurement
of the CCD orientation using the aluminum mask (Sec.~\ref{sec:X-method})
and using the quartz mask (Sec.~\ref{sec:optical-method}).
In section~\ref{sec:temp-dep} we describe the measurement of the inter--pixel distance temperature dependence.

\section{Set--up of the optical measurement} \label{sec:set-up}

\begin{figure}
\includegraphics[width=0.25\textwidth]{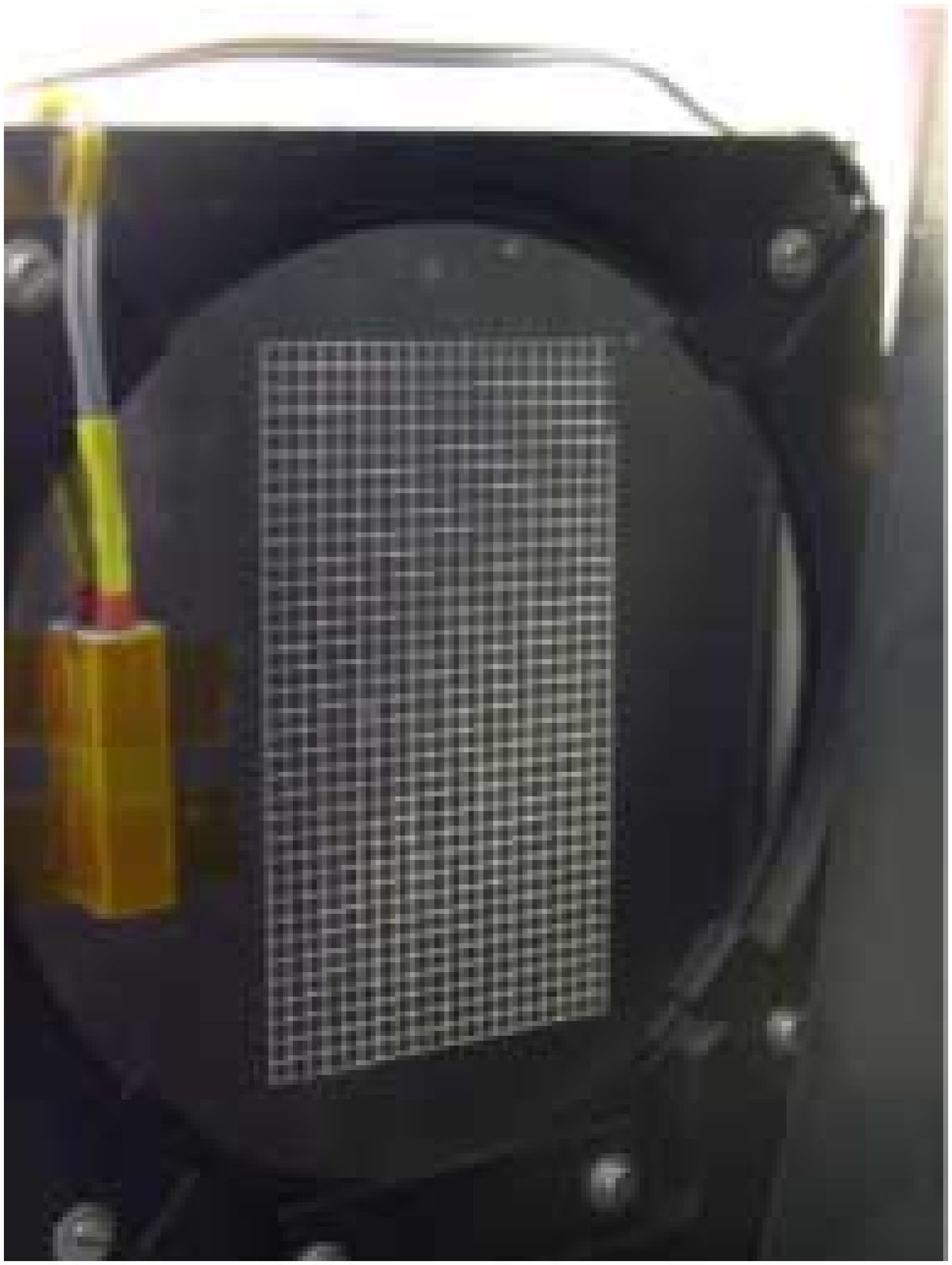} 
\caption{Quartz wafer illuminated by light. The spacing of the grating is 2\,mm both
         vertically and horizontally.}
\label{wafer}
\end{figure}

\begin{figure}
\includegraphics[width=0.5\textwidth]{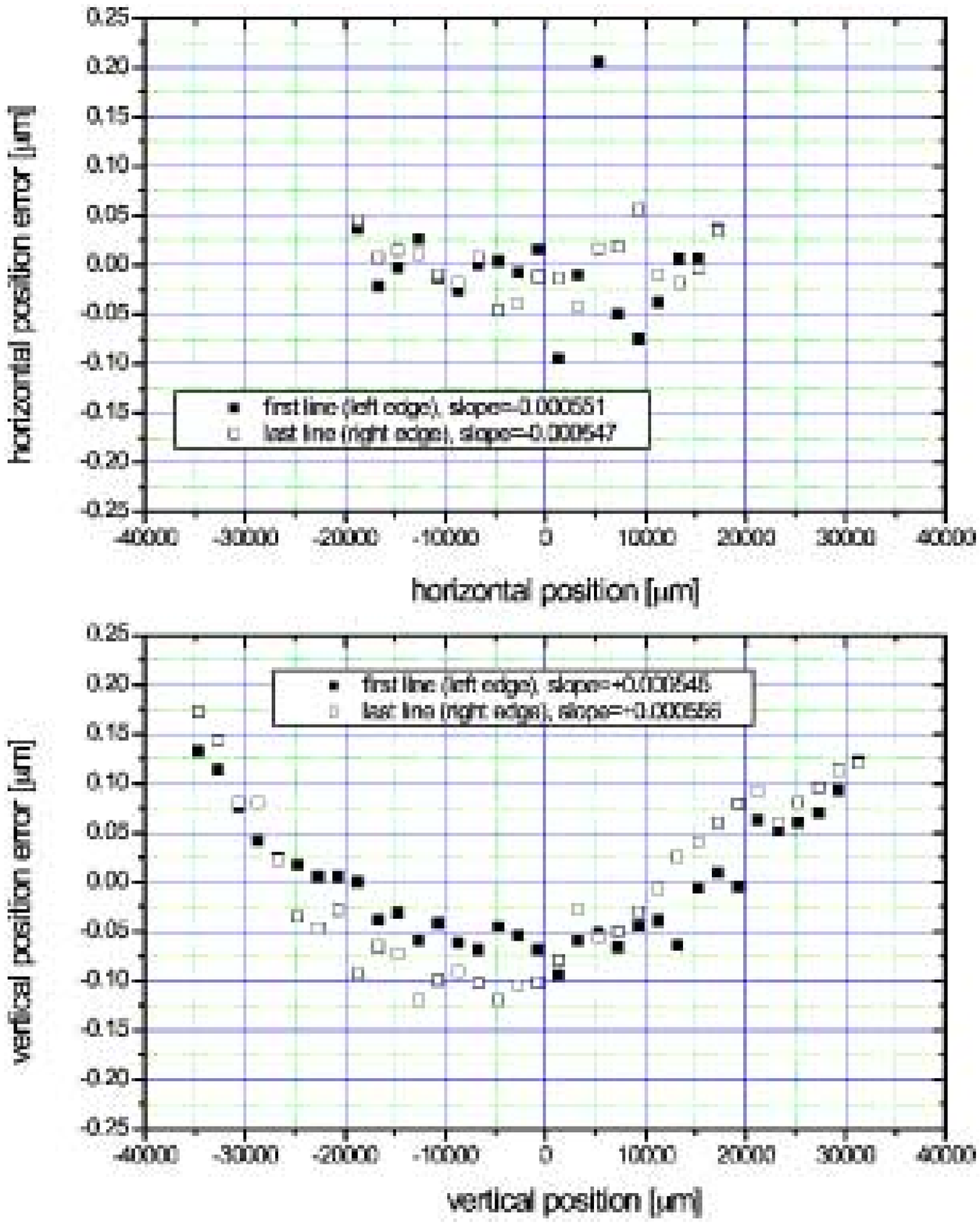} 
\caption{Linearity of the grating in horizontal direction (top) and vertical direction (bottom).}
\label{linearity}
\end{figure}

The quartz wafer is an optical cross grating  manufactured by  
the Laboratory of Micro- and Nanotechnology of the Paul Scherrer Institut. 
The grating is 40~mm
wide and 70 mm high. It is composed of vertical and horizontal lines of 
50~$\mu $m thickness separated from each other by 2~mm (Fig.\,\ref{wafer}).
The linearity of the lines is of order 0.05~$\mu$m in the horizontal 
direction. In the vertical direction, the lines become slightly parabolic with  
a maximum deviation of 0.15~$\mu$m from the average value (Fig.~\ref{linearity}).

\begin{figure}
\includegraphics[width=0.4\textwidth]{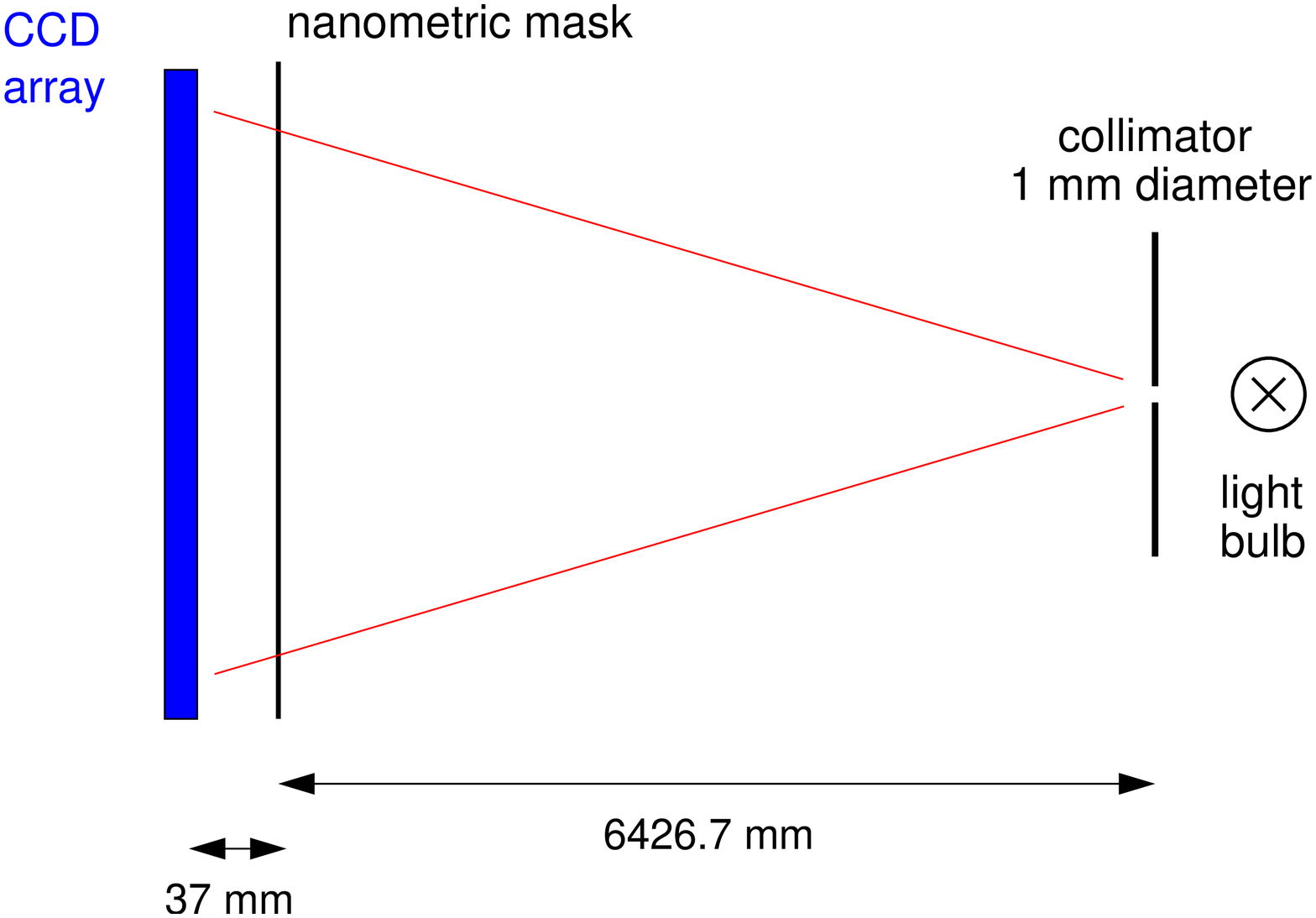}
\caption{Scheme of the experimental set--up.}
\label{setup}
\end{figure}

\begin{figure}
\includegraphics[height=0.4\textheight]{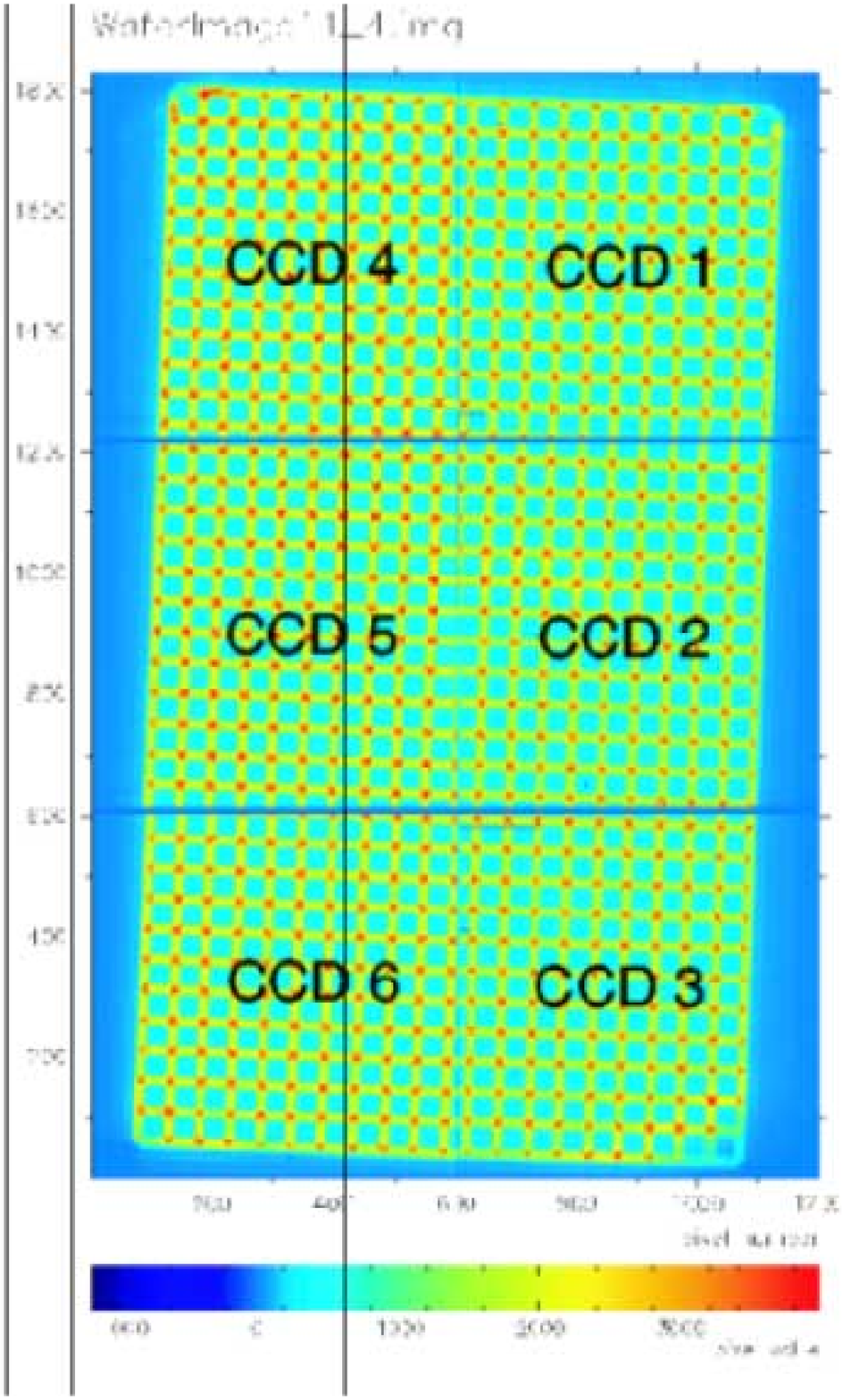}  
\caption{Image of the quartz wafer as seen without correcting for the relative 
               positions of the CCDs.}
\label{setup-bis}
\end{figure}

The wafer was positioned 37~mm in front of the CCD array. It was illuminated with short light 
pulses using a point--like light source, which was approximated by a collimator of one millimeter 
in diameter located in front of a light bulb at a distance of 6.43\,m from the CCDs 
to reduce parallax effects distorting the wafer image (Fig.~\ref{setup}-\ref{setup-bis}). 
The wafer temperature was monitored and remained at room temperature during the measurements.
The integration time per picture was 10~s with the bulb shining for 6~s for each selected  
temperature of the CCDs. The temperature was varied between $-105^\circ $C 
and $-40^\circ $C.

\section{Measurement of the average pixel distance}\label{sec:pxd}

For the determination of the pixel distance, a simultaneous linear fit of two adjacent lines
was performed under the constraint that the two lines are parallel. 

\begin{figure}
\includegraphics[width=0.3\textwidth]{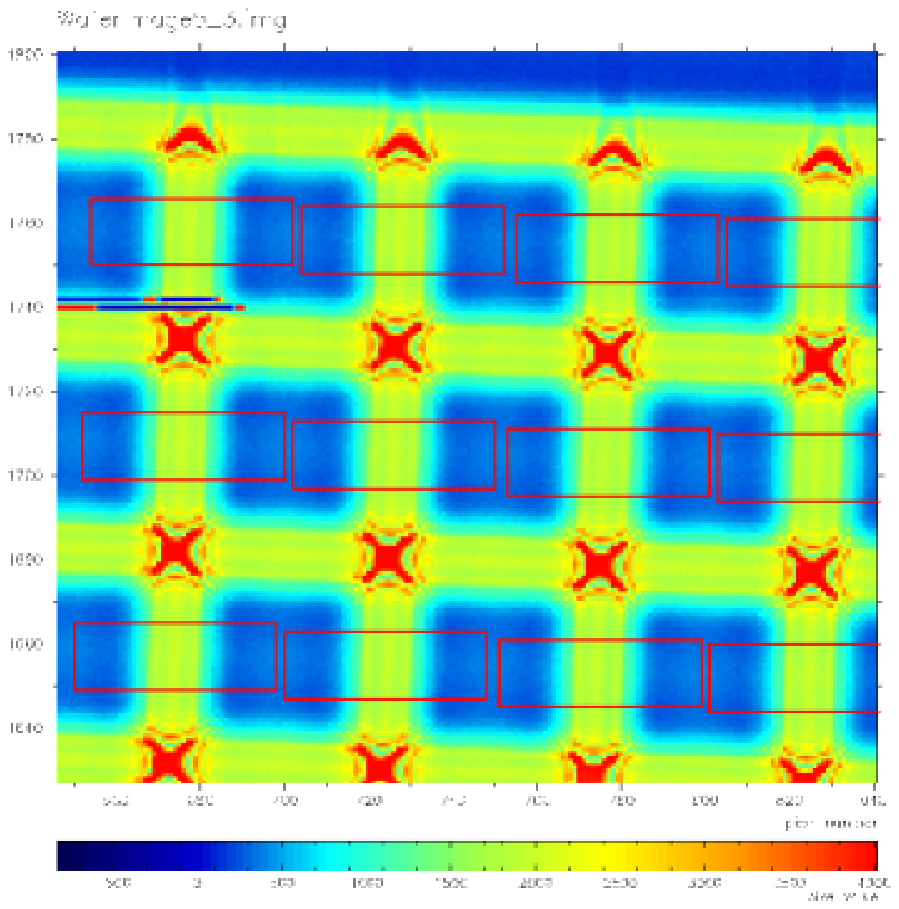}  
\caption{Selection of the line fitting zones  on the wafer image materialized by solid line rectangles.}
\label{zones}
\end{figure}

\begin{figure}
\includegraphics[width=0.4\textwidth]{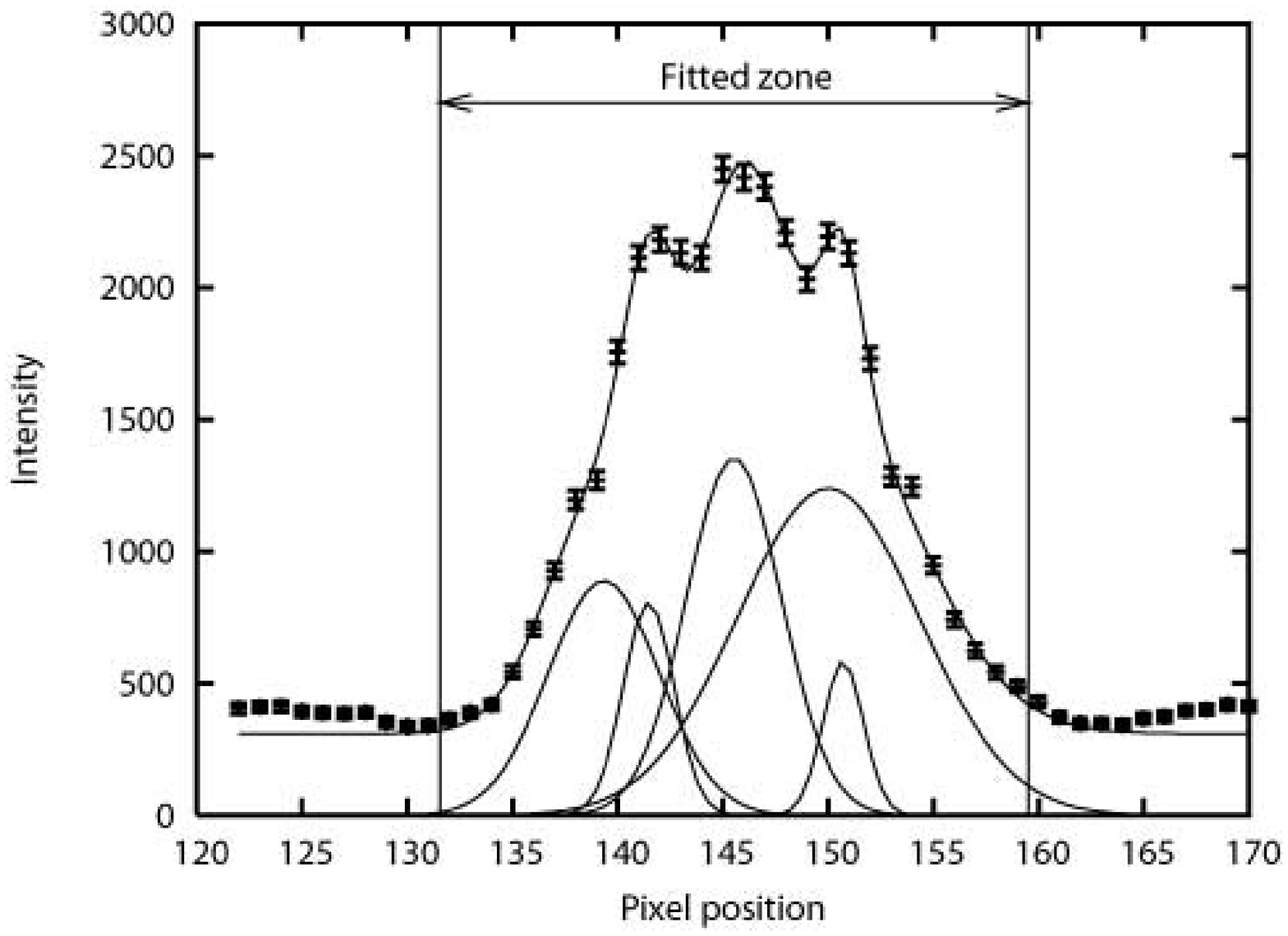} 
\caption{Intensity profile of one pixel row of a selected zone. 
         The line position is defined by using the 
          average of the three central profiles. The other two profiles, 
          normally characterized by a larger width, strongly depend on the background, i.~e., 
          on the illumination conditions of the selected zones.}
\label{zones-bis}
\end{figure}

After cutting out the crossing points, 
the diffraction pattern of the straight sections linking them (zones) was fitted to a superposition of 5 Gaussian 
profiles: central peak, first and second side maxima, and left and right backgrounds (Fig.~\ref{zones}-\ref{zones-bis}). 
The parabolic shape of the grating was taken into account in the analysis of the 
images recorded with the detector.

For the fit of two parallel lines we have to consider two sets of data at the same
time: $(x1_i, y1_i, \Delta y1_i)$ and $(x2_i, y2_i, \Delta y2_i)$, and the lines are described by the equations:
\begin{equation}
\begin{cases}
y1 = a1 + b\ x1 \\
y2 = a2 + b\ x2
\end{cases}.
\end{equation}
The best determination of the parameters $a1$, $a2$ and $b$ is obtained by
minimization of the $\chi^2$ merit function following the same procedure as described in Ref.~\cite{NumRec}.  
In this case, the $\chi^2$ merit function is:
\begin{multline}
\chi^2(a1,a2,b) = \sum^{N_1}_{i=1}\left(\frac{y1_i- a1 - b\ x1_i}{\Delta y1_i}\right)^2 + \\
\sum^{N_2}_{i=1}\left(\frac{y2_i- a2 - b\ x2_i}{\Delta y2_i}\right)^2. \label{eq:chi2}
\end{multline}
Considering two parallel lines that are at a distance $L$ (in $\mu$m) on the CCD,
the average pixel distance is obtained from the formula:
\begin{equation}
\text{pixel dist.} = \frac{L}{|a1 -a2| \cos (\arctan b)} = L \frac{\sqrt{1+b^2}}{|a1 -a2|}.
\end{equation}
The presence of the cosine term takes into account the fact that
the lines are generally not parallel to the CCD edge.
The detailed formulas for the $\chi^2$ minimization are presented in Appendix~\ref{app:2lines_fit}.

\begin{figure}
\includegraphics[width=0.4\textwidth]{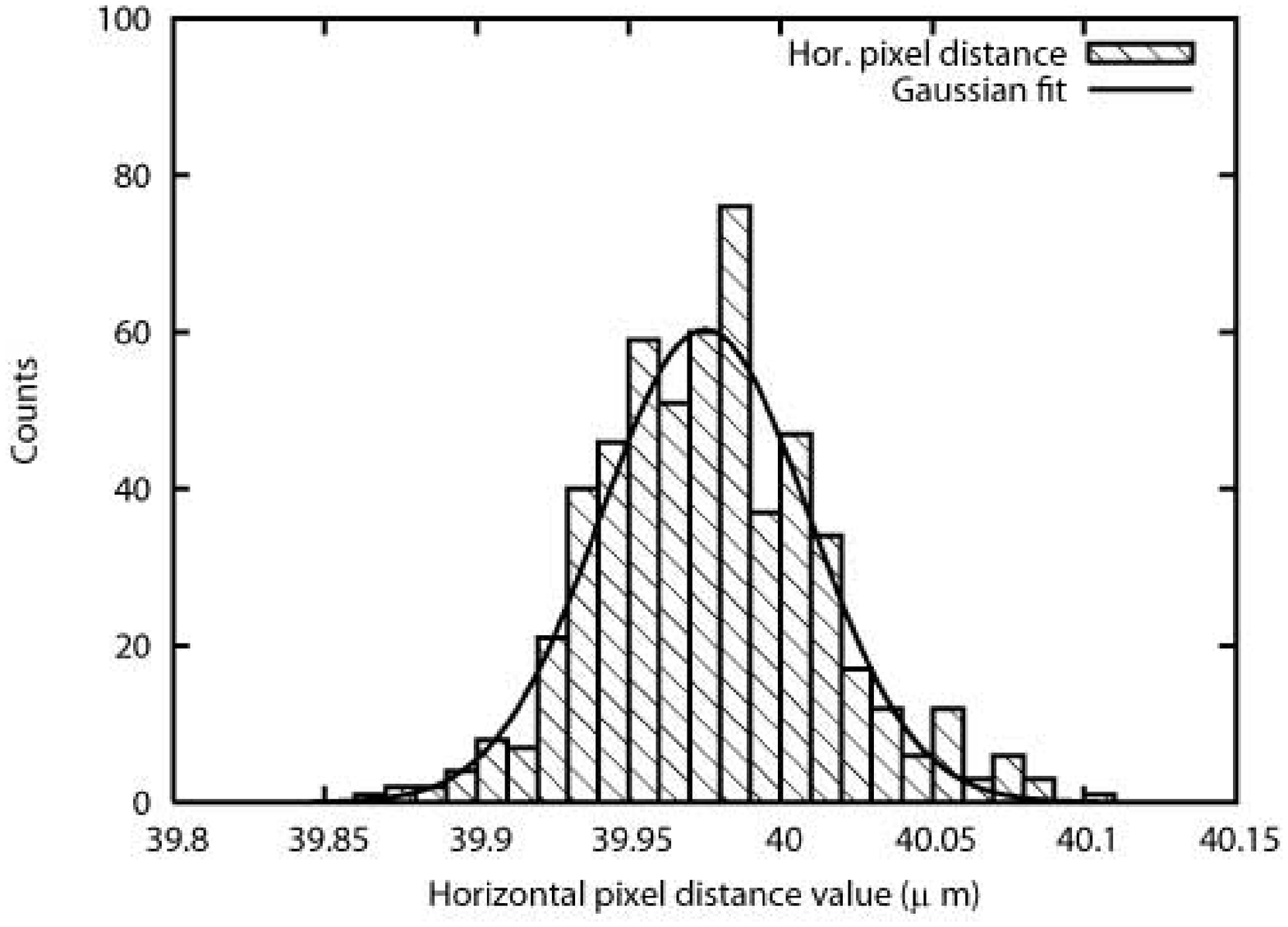} 
\caption{Distribution of the horizontal pixel distance 
          in CCD\,3 as obtained from pairs of selected zones.}
\label{pxdx_dist}
\end{figure}

\begin{figure}
\includegraphics[width=0.4\textwidth]{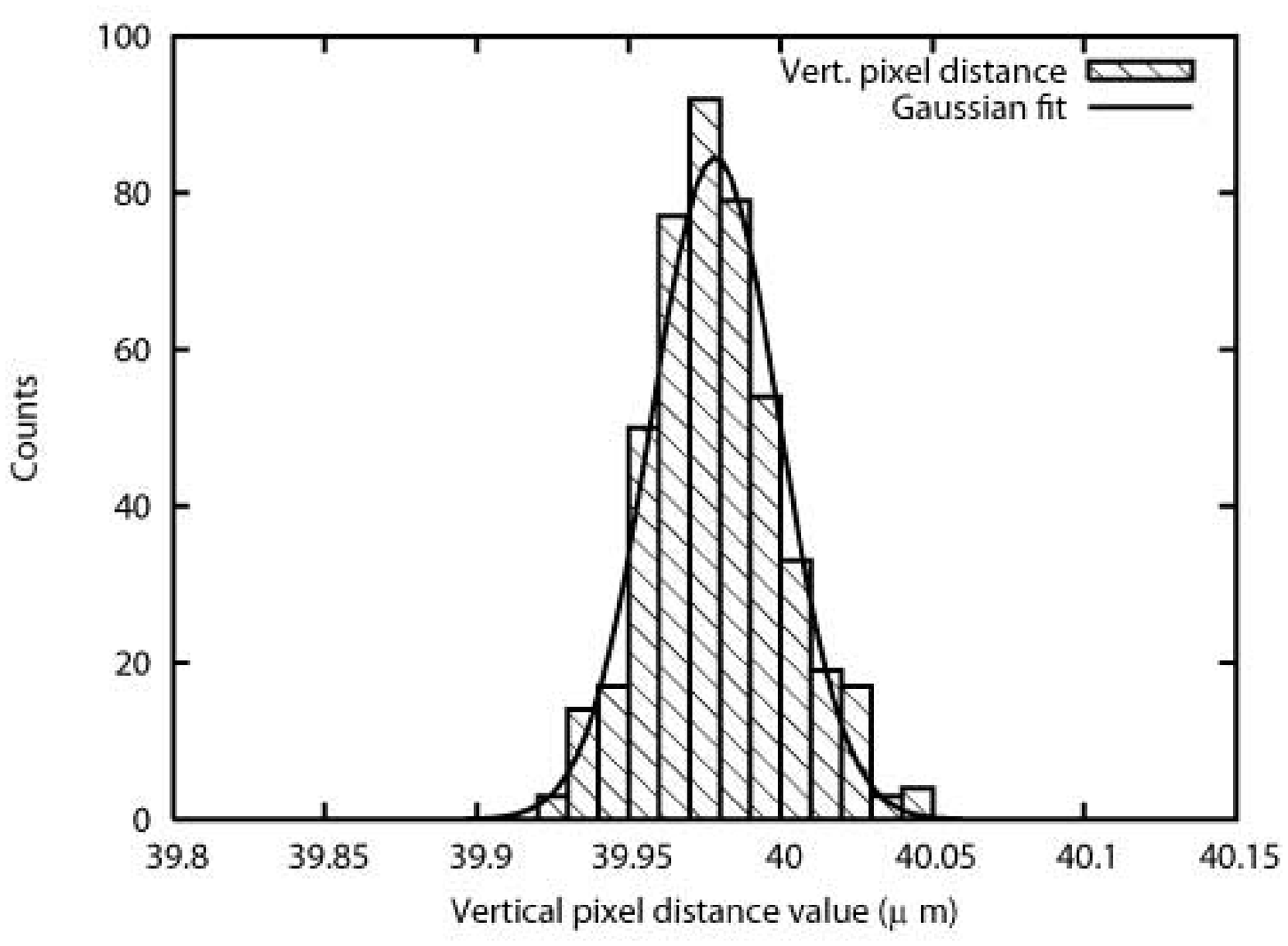}
\caption{Distribution of the vertical pixel distance 
         in CCD\,3 as obtained from pairs of selected zones.}
\label{pxdy_dist}
\end{figure}

For each CCD, we obtained about 180 independent evaluations of the pixel distance from
straight sections of different line pairs.
The average value of the pixel distance was obtained by a Gaussian fit to the  
histogram obtained from individual values (Fig.~\ref{pxdx_dist}-\ref{pxdy_dist}).
Two series of images were available and  the
final value was calculated from the sum of the two distributions. 

It is interesting to observe that the vertical and horizontal distributions have different dispersions 
(Fig.~\ref{pxdx_dist}-\ref{pxdy_dist} and Table~\ref{tab:gauss-fitxy}).
The horizontal pixel distance distribution is characterized by a FWHM of 80~nm, compared to  
50~nm for the vertical one. Accordingly, the error on 
the Gaussian peak position for the vertical distance is half that  for the horizontal one
(0.9~nm and 1.8~nm, respectively). 
We have no clear-cut explanation for this difference. 
It is unlikely that this difference could come from the accuracy of the
mask fabrication. As seen from Fig.~\ref{linearity}, the line distances show similar fluctuations 
in the order of 0.05~$\mu$m for both directions
and they should produce a dispersion of about 0.05~$\mu$m / 50 = 1~nm
on the vertical and horizontal pixel distance (50 is the average number of pixels between two lines in
the wafer image).

The CCD devices were fabricated using a 0.5~$\mu$m technology, which means that
the uncertainty over the full size is 0.5\,$\mu$m (at room temperature). Such an inaccuracy 
could introduce an average difference of order 0.8~nm for the 
inter--pixel distance of various CCDs. 
This assumption was tested applying {\it Student's t-test} \cite{NumRec} 
to distributions from different CCDs.
The only significant difference in the obtained distributions comes from
CCD\,2 and CCD\,5. However, for these two CCDs
we observe a parasitic image of the mask superimposed on the normal one, probably due to a reflection between 
the detector and the mask itself. 
Therefore, the final value of the pixel distance is given by the 
weighted average of the individual CCD values excluding CCD\,2 and CCD\,5 
(Table~\ref{tab:gauss-fitxy}).

The overall precision of the quartz wafer is quoted to be $\pm 0.0001$~mm over the full width of 40~mm.
Hence, the uncertainty of the wafer grid contributes on average  
0.1\,$\mu$m / 1000 = 0.1~nm per pixel.
As horizontal and vertical pixel distances are in good agreement, a weighted average is 
calculated. 
Taking the wafer uncertainty of 0.1~nm into account, the average pixel distance reads
$39.9775 \pm 0.0005 \pm 0.0001\,\mu m$, where the nominal value is 40$\,\mu m$. 

\begin{table} \small
\caption{Results of a Gaussian fit to the horizontal and vertical pixel distance distribution.
         The fabrication accuracy of the quartz wafer contributes with additionally 0.1~nm to the average pixel distance.} 
\label{tab:gauss-fitxy}
\begin{tabular}{c c c c}   
\hline
CCD    &       Hor.  dist. ($\mu$m)       &       FWHM ($\mu$m)      &       $\chi^2$    \\
\hline                                                                                  
1       &       $39.9778 \pm       0.0018$  &       $0.0820  \pm       0.0035$  &       1.11    \\
2       &       $39.9743 \pm       0.0018$  &       $0.0810  \pm       0.0033$  &       1.26    \\
3       &       $39.9751 \pm       0.0018$  &       $0.0808  \pm       0.0033$  &       1.41    \\
4       &       $39.9753 \pm       0.0017$  &       $0.0808  \pm       0.0032$  &       1.16    \\
5       &       $39.9744 \pm       0.0017$  &       $0.0856  \pm       0.0031$  &       1.01    \\
6       &       $39.9777 \pm       0.0018$  &       $0.0913  \pm       0.0031$  &       1.20    \\
\hline
weighted average & $39.9764 \pm  0.0009$  & \multicolumn{2}{l}{without CCDs 2 and 5} \\
line fits \\
\hline
 CCD    &       Vert.  dist.   ($\mu$m)     &        FWHM ($\mu$m)    &       $\chi^2$     \\
\hline                                                                                  
1       &       $39.9766 \pm       0.0012$  &       $0.0504  \pm       0.0022$  &       1.05    \\
2       &       $39.9787 \pm       0.0008$  &       $0.0420  \pm       0.0014$  &       0.88    \\
3       &       $39.9785 \pm       0.0010$  &       $0.0496  \pm       0.0019$  &       0.68    \\
4       &       $39.9769 \pm       0.0009$  &       $0.0450  \pm       0.0016$  &       0.62    \\
5       &       $39.9781 \pm       0.0007$  &       $0.0472  \pm       0.0013$  &       0.52    \\
6       &       $39.9787 \pm       0.0007$  &       $0.0423  \pm       0.0014$  &       0.66    \\
\hline
weighted average & $39.9779 \pm   0.0004$  & \multicolumn{2}{l}{without CCDs 2 and 5} \\
line fits \\
\hline                                  
\end{tabular}
\end{table}

\section{Measurement of the relative orientation of the CCDs}\label{sec:CCD-orientation}

\subsection{X--ray method}\label{sec:X-method}

An aluminum mask was installed 37~mm in front of the CCD array; this mask has a slit pattern 
chosen to provide an unambiguous connection between all CCDs (Fig.\,\ref{Al_mask_PSI}).
The mask has a thickness of 1\,mm, the slits are wire eroded with a width of about 0.1\,mm and the 
linearity is about 50~$\mu m$ over the full height. 
The detector array, shielded by the mask, was irradiated with sulphur X--rays of 2.3\,keV produced with the help of 
an X--ray tube; this energy is low enough to keep charge splitting effects small \cite{Anagnostopoulos2003a}. 
The sulphur target was placed at about 4\,m  from the detector. A collimator with a diameter of 5\,mm was placed close
to the target to provide a point--like source. In total, about  600\,000 X-ray events were collected.

\begin{figure}
\includegraphics[width=0.4\textwidth]{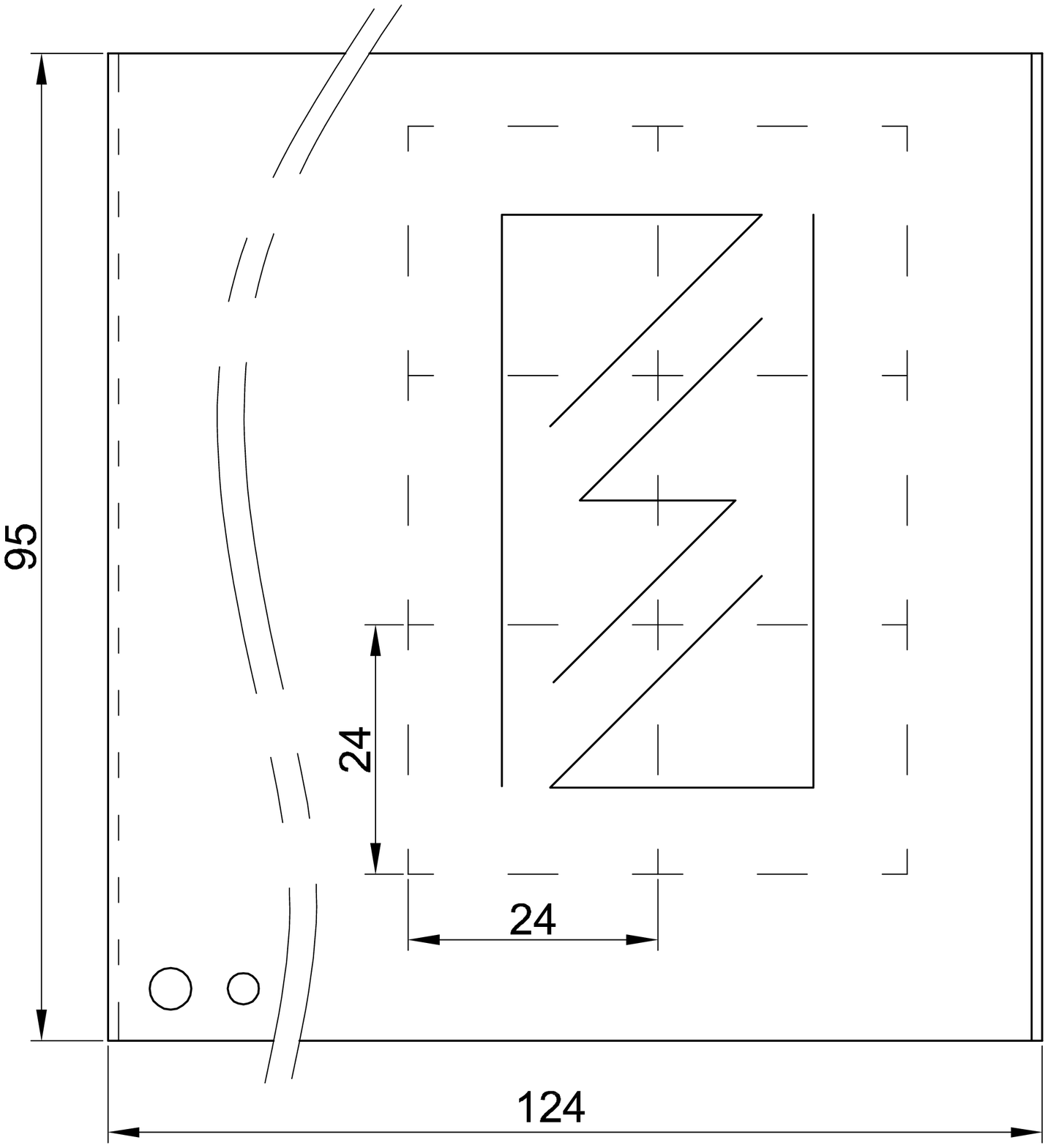}
\caption{Wire--eroded aluminum mask for the CCD alignment.}
\label{Al_mask_PSI}
\end{figure}

\begin{figure}
\includegraphics[width=0.4\textwidth]{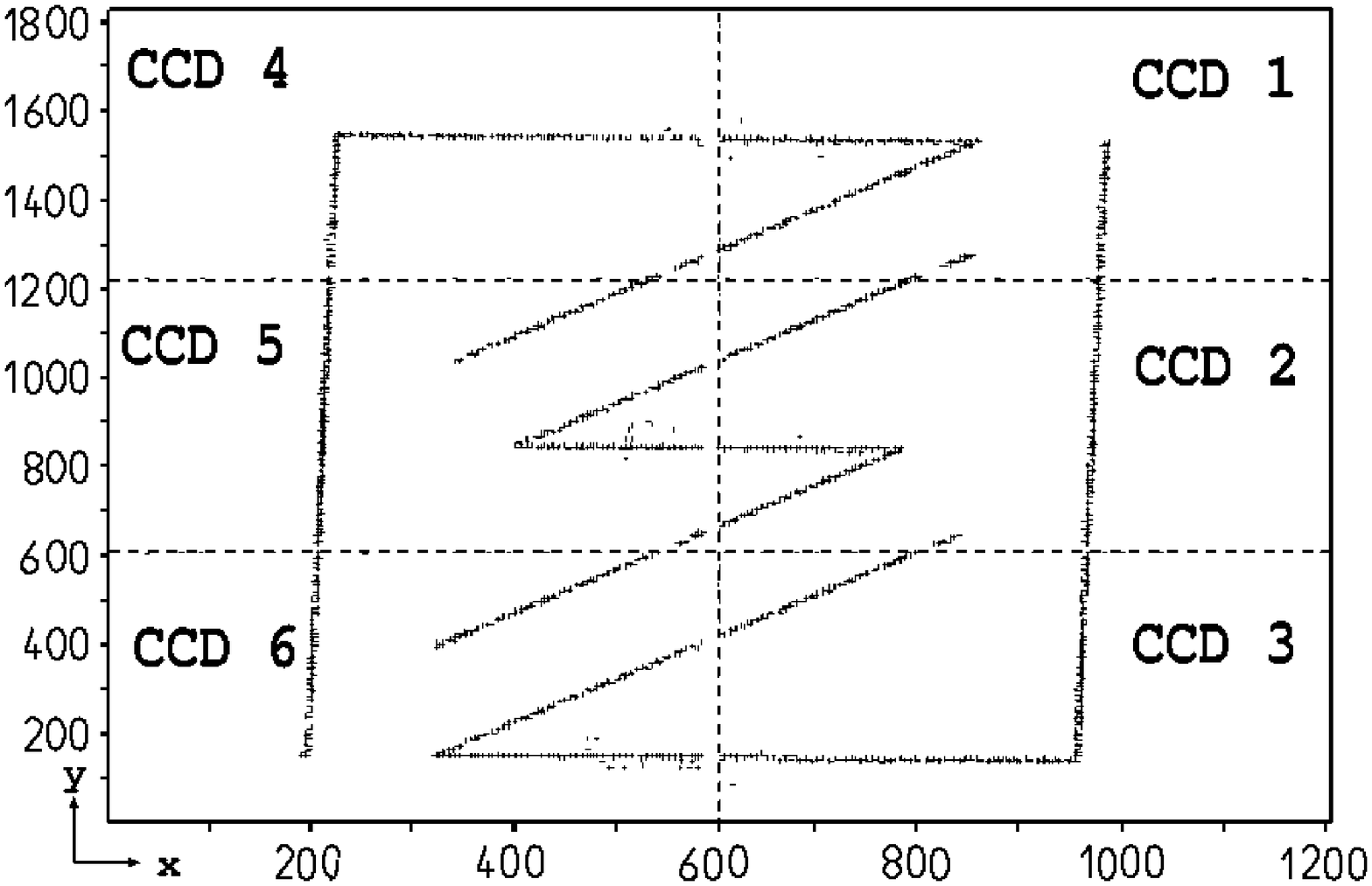}
\caption{Pattern produced by sulphur $K\alpha$ radiation excited by means of an X--ray tube.}
\label{Al_mask_PSI}
\end{figure}

The relative rotations of the CCDs are determined by performing linear fits to sections of the 
mask slit images. Because of the slit arrangement, CCD\,3 (CCD\,6 would be equivalent) is the best choice to serve 
as reference frame. In this case, the relative rotations of CCDs 1,2 and 6 are established directly. 
The values for CCD\,4 and CCD\,5 are the weighted average of results with CCD\,1 and CCD\,6 as intermediate 
steps.

The fit is done by calculating the center of gravity (COG) for each CCD row (or column for 
fitting a horizontal line) and then making a linear regression through them. The error of the COGs is 
based upon a rectangular distribution with a width equal to the width of the slits of the mask. With 
{\it N} as the number of events and {\it w} as the slit width, 
$\Delta_{COG}=\frac{w}{\sqrt{12} \cdot \sqrt{N}}$. A width $w$ of 4 pixels for the horizontal/vertical 
lines and 6 pixels for the diagonals is assumed.
From the inclinations (in mrad) of the mask slits relative to the perfect horizontal, vertical or 
diagonal (45$^\circ$), the rotations $\Delta \Theta$ of individual CCDs are calculated. Results 
(relative to CCD\,3) are given in Table\,\ref{CCD_align_sulphur}.

\begin{figure}[h]
\includegraphics[width=0.5\textwidth]{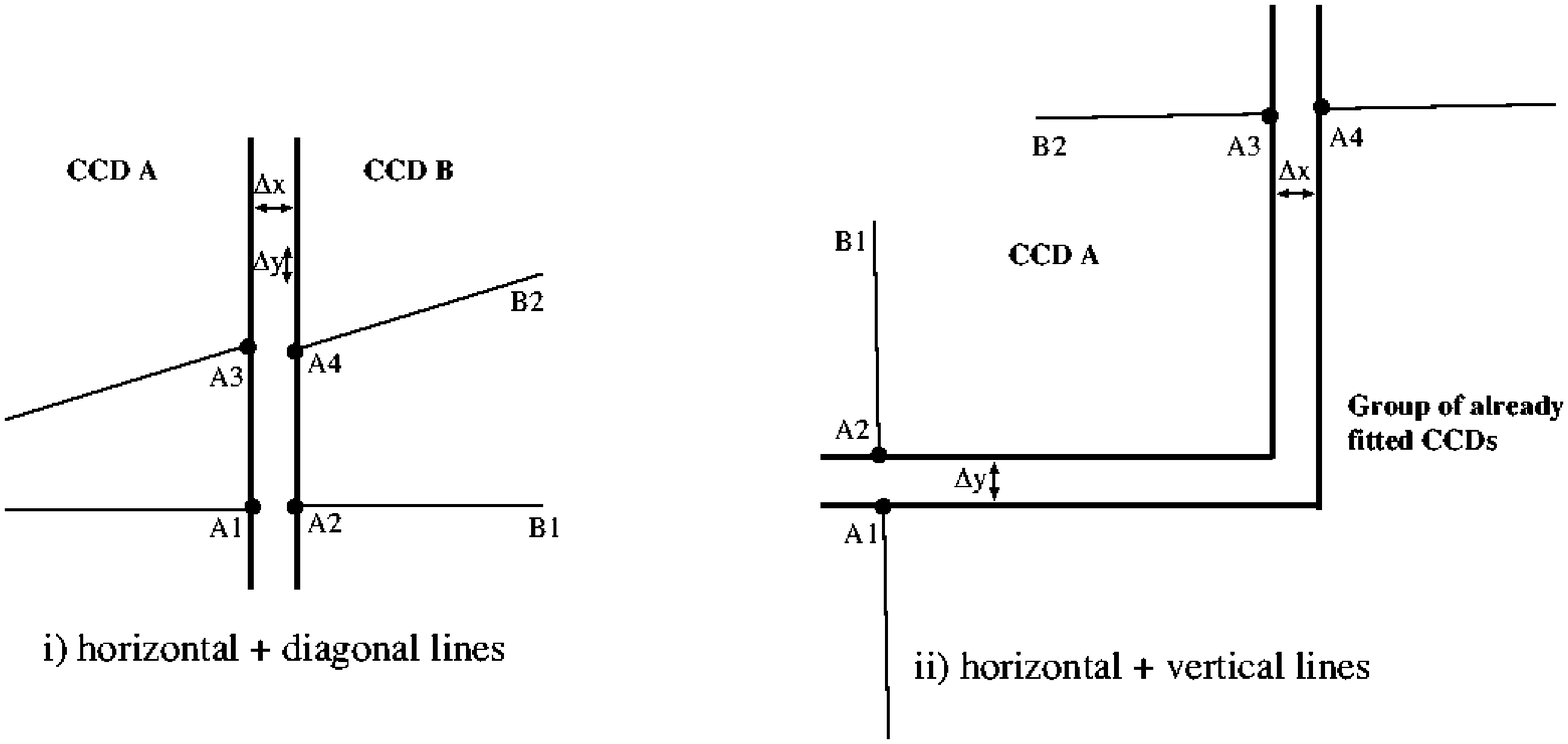}
\caption{Definition of crossing points for the determination of the relative offsets 
                of the CCDs.}
\label{offset}
\end{figure}

After the rotations have been determined and corrected for, the lines were 
fitted again to determine the crossing points of each slit with the CCD edge. The relative 
offsets $\Delta x$ and $\Delta y$ can be determined only if there are at least two lines crossing from one CCD to the 
other (Fig.\,\ref{offset}). With CCD\,3 as the starting point, the only other CCD fulfilling this 
condition is CCD\,6. The position of all other CCDs has to be calculated relative to  all 
CCDs shifted so far. The correct order for this is CCD\,2, then CCD\,5, CCD\,1 and CCD\,4. 

The correct values for the vertical offsets follow from the condition that {both} 
lines should continue from one CCD to the other (CCD\,A and CCD\,B in Fig.\,\ref{offset}). 
For case i) in Fig.\ \ref{offset}, one horizontal and one diagonal line:
\begin{equation}
\begin{cases} \label{eq:starting-eq}
A_1 + \Delta y + B_1 \cdot \Delta x = A_2 \\
A_3 + \Delta y + B_2 \cdot \Delta x = A_4
\end{cases},
\end{equation}
where $A_i$ are the y-coordinate of the crossing point between the lines of
equation $y = B_i \cdot x +$(constant)  and the CCD edge.
From this, one derives:
\begin{equation}
\Delta x = \frac{(A_2 - A_4) - (A_1 - A_3)}{B_1 - B_2}, 
\end{equation}
and the associate error is:
\begin{multline}
\delta (\Delta x) = \Big[ \frac{(\delta A_1)^2+(\delta A_2)^2+(\delta A_3)^2+(\delta A_4)^2}{(B_1 - B_2)^2} + \\
\frac{((\delta B_1)^2+(\delta B_2)^2) \cdot (A_2 - A_4 - A_1 + A_3)^2}{(B_1 - B_2)^4} \Big]^{1/2}
\end{multline}
For case ii), one horizontal and one vertical line,
\begin{equation}
\begin{cases}
A_1 + \Delta x + B_1 \cdot \Delta y = A_2\\
A_3 + \Delta y + B_2 \cdot \Delta x = A_4
\end{cases}
\end{equation}
(note that $B_1$ is defined 
as $x = B_1 \cdot y + $(constant) ). Here, the equations are:
\begin{equation}
\Delta x = \frac{A_1 - A_2 - B_1(A_3 - A_4)}{B_1 \cdot B_2 - 1},
\end{equation}
\begin{multline}
(\delta (\Delta x))^2 = \frac{(\delta A_1)^2 + (\delta A_2)^2 + (\delta B_1)^2((\delta A_3)^2 + (\delta A_4)^2)}{(B_1 \cdot B_2 - 1)^2}\\
+ (\delta B_1)^2 \left( \frac{A_4 - A_3}{B_1 \cdot B_2 - 1} - \frac{B_2(A_1 - A_2 - B_1(A_3 - A_4))}{(B_1 \cdot B_2 - 1)^2} \right)^2 \\
+  (\delta B_2)^2 \left( \frac{B_1(A_1 - A_2 - B_1(A_3 - A_4))}{(B_1 \cdot B_2 - 1)^2} \right)^2.
\end{multline}

Values for $\Delta y$ are derived by inserting $\Delta x$ in either of the starting equations Eq.~\eqref{eq:starting-eq}. The final 
horizontal and vertical displacements (which depend on the previously determined set of rotations) 
are given in Tab.\,\ref{CCD_align_sulphur}.

\begin{figure}[h]
\includegraphics[width=0.4\textwidth]{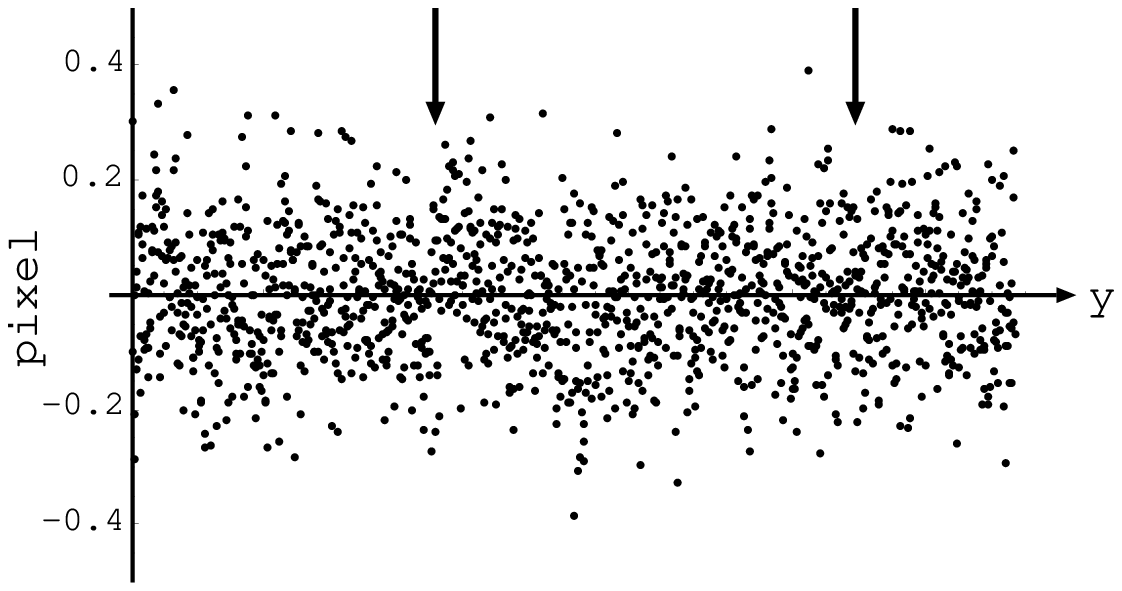}
\caption{Centers of gravity of the right vertical slit of the wire--eroded aluminum mask. 
                Arrows indicate the two largest irregularities.}
\label{cogplot}
\end{figure}

The analysis of the mask data assumes that the slits on the mask are perfectly straight; the 
given uncertainties are then purely statistical. However, a detailed study of the vertical slit to the 
right (on CCD\,1 to CCD\,3) shows that the mechanical irregularities of the mask are big enough to be 
noticeable. Fig.\,\ref{cogplot} shows the centers of gravity calculated for this slit subtracted 
from the fit through these points. Both the sudden jump (left arrow) and the inclination change 
(right arrow) are substructures on a scale of roughly 1/10th of a pixel (4\,$\mu$m). This fits well 
with the mechanical accuracy of 5\,$\mu$m quoted for the mask slits. Consequently, a
further improvement in accuracy is not limited by statistics, but by the mechanical precision 
of the mask itself. More details may be found in\,\cite{Hennebach2004}.

\begin{table}
\small
\caption{CCD position corrections (relative to CCD\,3) from the mask measurement 
         using sulphur fluorescence radiation.}
\label{CCD_align_sulphur}
\begin{tabular}{l r r r}
\hline
CCD    &       $\Delta x$ (pixels)                     &       $\Delta y$ (pixels)                     &       $\Delta \Theta$  (mrad)          \\
\hline
CCD3-CCD1       & $     -2.818  \pm     0.022   $ & $   22.264  \pm     0.077   $ & $   0.197   \pm     0.078    $ \\
CCD3-CCD2       & $     -1.049  \pm     0.015   $ & $   10.901  \pm     0.085   $ & $   0.522   \pm     0.062    $ \\
CCD3-CCD3       & $     0.000   \pm     0.000   $ & $   0.000   \pm     0.000   $ & $   0.000   \pm     0.000    $ \\
CCD3-CCD4       & $     -14.347 \pm     0.046   $ & $   20.808  \pm     0.075   $ & $   1.577   \pm     0.084    $ \\
CCD3-CCD5       & $     -14.597 \pm     0.043   $ & $   12.265  \pm     0.064   $ & $   2.940   \pm     0.109    $ \\
CCD3-CCD6       & $     -16.487 \pm     0.040   $ & $   1.173   \pm     0.052   $ & $   6.328   \pm     0.101    $ \\
\hline
\end{tabular}
\end{table}

\subsection{Optical method}\label{sec:optical-method}

\begin{figure}
\includegraphics[width=0.5\textwidth]{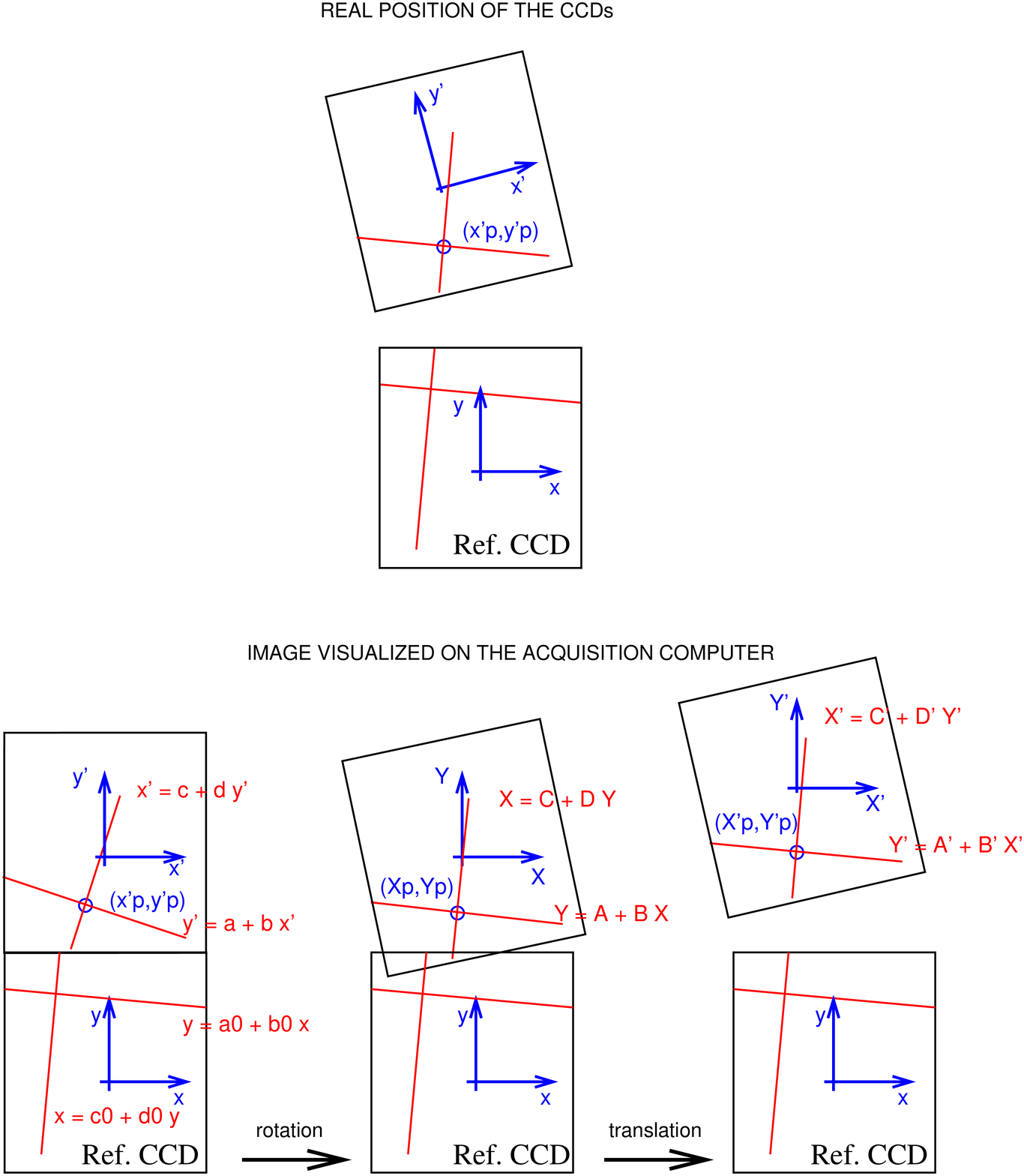} 
\caption{Scheme of the transformation  used in obtaining the orientation and shift 
         between CCDs. In the top part, the real position of the CCDs is shown together 
         with one crossing of the quartz grid. 
         In the lower part, the transformation from the individual CCD coordinates (left) to the real 
         relative position with respect to the reference CCD is displayed. The 
         rotation is first performed (middle) and then the shift is adjusted from the  known
         geometry of the grid (right).}
\label{ccd-rot}
\end{figure}

By using the nanometric quartz wafer, the precision for the CCD offsets was improved 
beyond 1/20 of the pixel width of 40\,$\mu$m, which was envisaged for measuring the 
charged pion mass. The knowledge of the line positions on the wafer allows one  
to infer the relative position between pairs of CCDs from the image. 
As for  X--rays, the image, when visualized without position and rotation correction, 
shows discontinuities at the boundaries of adjacent CCDs: lines are not parallel and
a part of the mask image is missing due to the spatial separation of the CCDs 
(Fig.~\ref{ccd-rot} bottom--left). Again, one CCD has to be chosen as a reference.

The unambiguous calculation of relative horizontal and vertical shift 
($\Delta x$ and $\Delta y$) and rotation ($\Delta \Theta$) of two CCDs requires
the information coming from at least one pair of perpendicularly crossing lines per CCD.
Using the line parameters, it is possible to build a function depending upon 
$\Delta x$, $\Delta y$ and $\Delta \Theta$, which is minimal when the shift and rotation values are 
optimal.
The idea is to compare the coordinates of a crossing point using the reference 
frame of the reference CCD ($x_p$, $y_p$) and of the selected CCD ($x'_p$, $y'_p$).
The values of $\Delta x$ and $\Delta y$ are unequivocally determined by first applying a rotation of the
coordinate system of the selected CCD around the CCD center. 
The value of the rotation angle $\Delta \Theta$ 
is chosen to have the lines parallel to the ones of the reference CCD 
(Fig.~\ref{ccd-rot} bottom-middle side).
In this new frame, the coordinates $(X_p,Y_p)$ of the crossing point depend 
on the line parameters and on the value of $\Delta \Theta$.
The differences $X_p - x_p$ and $Y_p - y_p$ provide exactly the shift values $\Delta x$ and $\Delta y$.
A function $F$ may be defined as:
\begin{equation}
\label{defF}
F(\Delta x, \Delta y, \Delta \Theta) = (X_p - x_p - \Delta x)^2 + (Y_p - y_p - \Delta y)^2
\end{equation}

In the ideal case, $F = 0$,  the values of $\Delta x$, $\Delta y$ and $\Delta \Theta $ are the correct ones.
In reality we assume that, for a selected set of lines, 
the best estimate of $\Delta x, \Delta y$ and $\Delta \Theta $ is found when $F$ is minimal.
The full expression used for F is given in appendix~\ref{app:F}.

\begin{figure}
\includegraphics[width=0.4\textwidth]{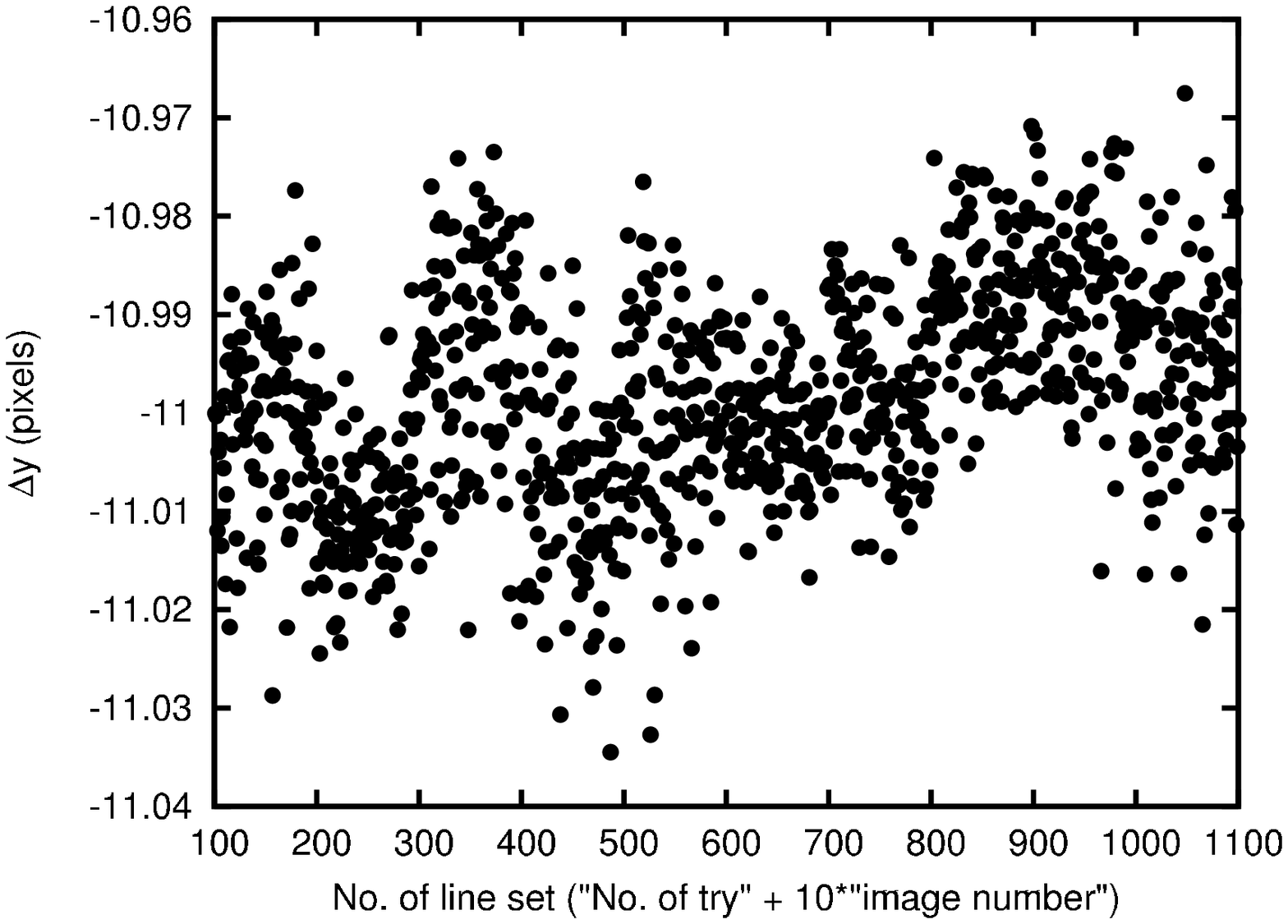} 
\caption{Distribution of the relative shift $\Delta y$ (gap) for CCD\,3 relative to 
         reference CCD\,2 for various crossing points.
         Each point corresponds to a value of $\Delta y$ obtained for a set of line pairs. For
         each of the 10 images, 100 sets of line pairs have been randomly chosen.
         The slope with time (corresponding approximately to the ``No. of line set'' axis) 
         may be due to the CCD array not reaching the thermal equilibrium.}
\label{gapy}
\end{figure}

\begin{figure}
\includegraphics[width=0.4\textwidth]{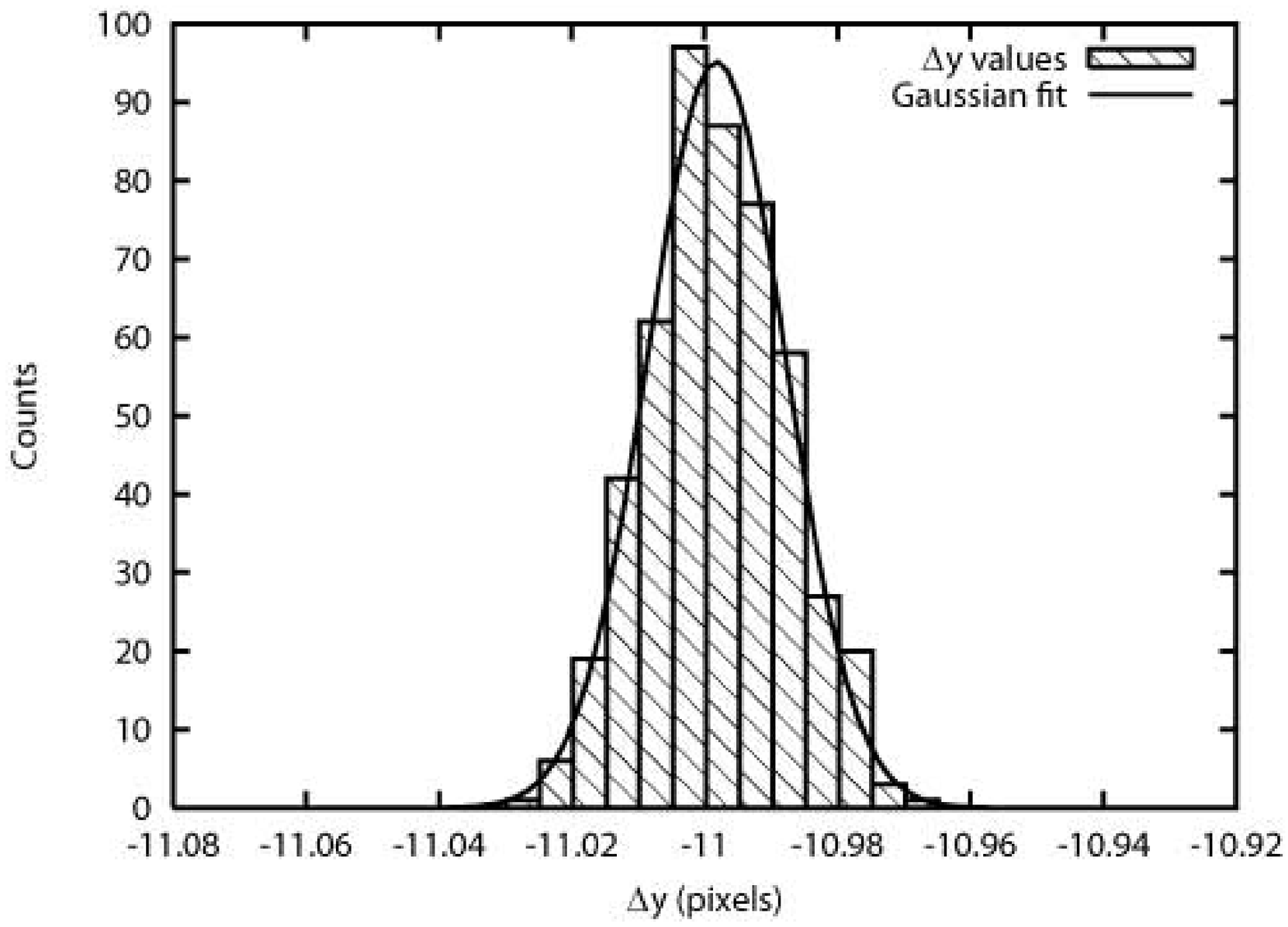}
\caption{Projection of the $\Delta y$ distribution. A Gaussian fit yields 
         the most likely value and an estimate for the uncertainty of $\Delta y$.}
\label{gapy}
\end{figure}

A whole set of values was obtained by randomly selecting line pairs.
For different choices of line pairs, different values are obtained for the position parameters.
Hence, the final values of $\Delta x$, $\Delta y$ and $\Delta \Theta$ are given again  
by a Gaussian fit to the distribution of the individual values. 
The accuracy of this method can be increased by forcing the simultaneous minimization 
of coordinate differences for several crossing points instead of only one.
Here, four crossing points and a set of 100 different choices of line pairs were used.
In this case the function $F$ reads
\begin{equation}
F(\Delta x, \Delta y, \Delta \Theta) = \sum^{4}_{i = 1}(X_p^i - x_p^i - \Delta x)^2 + (Y_p^i - y_p^i - \Delta y)^2
\end{equation}
where $i$  = 1 to 4 corresponds to the crossing point number arbitrarily ordered.
Figure~\ref{gapy} shows the distribution data for $\Delta y$ obtained for the full set of line pairs.

The final result for the relative CCD positions was obtained from three series of 10 images each: 
two at $-100^\circ $C and one at $-105^\circ $C.
The precision for each series is around 0.001 pixels for $\Delta x$ and $\Delta y$, 
and $3~\mu$rad  for the relative rotation $\Delta \Theta $, and it can be reduced using a function $F$ with more crossing points.
The systematic errors were estimated by comparing the results from the 
three series of data acquisition. However, the differences between values from different series are of order 
0.01\,--~0.03 pixels for $\Delta x$ and $\Delta y$, and $50\,\mu rad$ for $\Delta \Theta $. 
This large spread, compared to the precision of each series, has two possible explanations: 
differences of the wafer illumination condition (affecting the line fit), 
or a mechanical change of the CCD array position during warming up and cooling of the 
detector. 
The second hypothesis is more likely, because only small differences were observed 
between the series at $-105^\circ $C and the first series at $-100^\circ $C, where no warming up
between the two measurements was performed. In contrast, before the second series 
at$-100^\circ $C, the detector was at room temperature for a short period.
This hypothesis is also confirmed by the observation of a small change in time of the 
$\Delta y$ values in Fig.~\ref{gapy}, where a significant change is observed between points obtained from
different images. These differences could be attributed to a mechanical change in time
due to the not yet attained thermal equilibrium of the CCD array during the measurement.

For each CCD, the final position and rotation parameters are calculated as the average of the 
three series (Table~\ref{tab:ccd_position}). 
The systematic effect from the temperature difference 
of the image series is negligibly small compared to the spread of values.
The systematic error is estimated using the standard deviation
formula for a set of values. For CCD\,4, only one series of measurements was available. In this
case, the largest value of all other CCDs was chosen.

The fabrication of the grating introduces a systematic error due to the slightly parabolic
shape of the vertical lines (Fig.\,\ref{linearity}). The error is estimated to be of order of
9\,$\mu$rad for $\Delta \Theta$ and 0.009~pixels for $\Delta x$ for CCD\,1, CCD\,3, CCD\,4 and 
CCD\,6, which is negligible compared with other systematic errors.

\begin{table} \small
\caption{CCD relative position and orientation with CCD\,2 as reference. The orientation of CCD\,2 relative to 
         itself provides a check of the validity of the measurement method.} 
\label{tab:ccd_position}
\begin{tabular}{l r r r }
\hline
CCD     &       $\Delta x$ (pixels)                     &       $\Delta y$ (pixels)                     &       $\Delta \Theta$ (mrad)          \\
\hline
CCD2-CCD1       & $     -1.251  \pm     0.029   $ & $   11.404  \pm     0.023   $ & $   -0.587  \pm     0.035    $ \\
CCD2-CCD2       & $     0.000   \pm     0.000   $ & $   0.000   \pm     0.000   $ & $   -0.002  \pm     0.003    $ \\
CCD2-CCD3       & $     0.509   \pm     0.012   $ & $   -11.021 \pm     0.021   $ & $   -0.677  \pm     0.074    $ \\
CCD2-CCD4       & $     -12.850 \pm     0.041   $ & $   10.279  \pm     0.023   $ & $   0.801   \pm     0.130    $ \\
CCD2-CCD5       & $     -13.579 \pm     0.009   $ & $   1.738   \pm     0.016   $ & $   2.233   \pm     0.130    $ \\
CCD2-CCD6       & $     -15.963 \pm     0.041   $ & $   -9.435  \pm     0.021   $ & $   5.530   \pm     0.011    $ \\

\hline
\end{tabular}
\end{table}

The values presented in Table~\ref{tab:ccd_position} are in very good agreement with the 
results obtained using the aluminum mask, taking into account the different reference CCD.
As an example, for the $\Delta x$ shift between 
CCD\,5 and CCD\,2 we obtain $-13.548 \pm 0.045$~pixels with the X-ray method, 
and $-13.579 \pm 0.009$~pixels with the optical method.   

\section{Temperature dependence of the pixel distance} \label{sec:temp-dep}
For the determination of the temperature dependence, images between  
$-105^\circ $C and $-40^\circ $C were acquired. For each condition the same analysis method as
described in Sec.~\ref{sec:pxd} was applied.
As expected, the pixel distance increases with increasing temperature except for the vertical 
pixel distance at $-40^\circ $C (Table~\ref{tab:pixel_temp}).
This effect may be caused by the high CCD read--noise level at this temperature.
The values obtained  at $-40^\circ $C have been ejected for the measurement of the temperature dependence. 

\begin{table}
\caption{Pixel distance values at different detector temperatures.}
\label{tab:pixel_temp}
\begin{tabular}{r r r }
\hline
Temp. ($^\circ $C)      &       Hor. pixel dist. ($\mu$m)                      &       Vert. pixel dist. ($\mu$m)                     \\
\hline
-105    & $     39.9796 \pm     0.0014  $ & $   39.9779 \pm     0.0006  $ \\
-100    & $     39.9764 \pm     0.0009  $ & $   39.9779 \pm     0.0004  $ \\
-80     & $     39.9796 \pm     0.0020  $ & $   39.9794 \pm     0.0006  $ \\
-60     & $     39.9827 \pm     0.0017  $ & $   39.9800 \pm     0.0006  $ \\
-40     & $     39.9837 \pm     0.0013  $ & $   39.9762 \pm     0.0010  $ \\
\hline                                      
\end{tabular}
\end{table}

The average of the thermal expansion coefficient is obtained by a simple linear extrapolation 
of the data between $-105^\circ $C and $-60^\circ $C. The results are:
 $(2.8 \pm 1.0 )\cdot 10^{-6} K^{-1}$ for the horizontal distance
and $(1.3 \pm 0.4 )\cdot 10^{-6} K^{-1}$ for the vertical distance.
These values are in the range of the thermal expansion coefficient  
of silicon, the CCD substrate material, and INVAR, the metallic support material
 for the temperatures considered: literature values are $0.8-1.6  \cdot 10^{-6} K^{-1}$ for 
silicon \cite{Lyon1977} and $1-2  \cdot 10^{-6} K^{-1}$ for INVAR \cite{Beranger}.

\section{Conclusion}
We have demonstrated that the average inter--pixel distance of a CCD detector under operating conditions 
can be determined to an accuracy of 15~ppm. We obtain 
$39.9775 \pm 0.0006\,\mu$m for the average pixel distance at a temperature of $-100^\circ $C, 
which deviates significantly from the nominal value of 40\,$\mu$m.
Also, the temperature dependence of the inter--pixel distance 
was studied and successfully compared to values found in the literature. 
The relative rotations and positions of the individual CCD devices of a $2 \times 3$ array have been measured to a 
precision of about $50~\mu$rad and $0.02$ pixel, respectively. The X--ray method was limited by 
the quality of the aluminum mask, i.\,e., by the accuracy of wire--eroding machine. With 
the nanometric quartz wafer no limitation occurs from the 
accuracy of the mask. The principal difficulty encountered in that case, is the proper description of the diffraction 
pattern and in particular the control of the illumination. The 
accuracy achieved by this method fully satisfies the requirements of a recent attempt to measure 
the charged pion mass to about 1.5~ppm. 
The X-ray method and the optical method can be used for any CCD camera sensitive to X-ray and/or
 visible light radiation.

\section*{Acknowledgments}
Partial travel support for this experiment has been provided by the ``Germaine de Sta{\"e}l'' French exchange program. 

\appendix 

\section{Formulas for fitting with a pair of parallel lines}\label{app:2lines_fit}
In this appendix, we present mathematical formulas for linear fitting with a pair of parallel lines,
i.e. for the minimization of the $\chi^2$ merit function defined in Eq.~\eqref{eq:chi2}.

%
%

$\chi^2$ is minimized when its derivatives with respect to $a1$, $a2$, and $b$ vanish:
\begin{equation}
\begin{cases}
0 = \cfrac{\partial \chi^2}{\partial a1} = -2 \sum^{N_1}_{i=1} \cfrac{y1_i- a1 - b\ x1_i}{\Delta y1_i^2} \\
0 = \cfrac{\partial \chi^2}{\partial a2} = -2 \sum^{N_2}_{i=1} \cfrac{y2_i- a2 - b\ x2_i}{\Delta y2_i^2} \\ 
0 = \cfrac{\partial \chi^2}{\partial b} = -2 \Big(  \sum^{N_1}_{i=1} \cfrac{x1_i(y1_i- a1 - b\ x1_i)}{\Delta y1_i^2} + \\
\qquad \qquad \sum^{N_2}_{i=1} \cfrac{x2_i(y2_i- a2 - b\ x2_i)}{\Delta y2_i^2} \Big) \label{eq:system}
\end{cases}.
\end{equation}
These conditions can be rewritten in a convenient form if we define the following sum:
\begin{align}
& S1 = \sum^{N_1}_{i=1} \cfrac{1}{\Delta y1_i^2},  & S1_x =  \sum^{N_1}_{i=1} \cfrac{x1_i}{\Delta y1_i^2},  \\
& S1_y = \sum^{N_1}_{i=1} \cfrac{y1_i}{\Delta y1_i^2},\\
& S1_{xx} = \sum^{N_1}_{i=1} \cfrac{x1^2_i}{\Delta y1_i^2}, & S1_{xy} =  \sum^{N_1}_{i=1} \cfrac{x1_i\ y1_i}{\Delta y1_i^2},  &\\
& S2 = \sum^{N_2}_{i=1} \cfrac{1}{\Delta y2_i^2},  & S2_x =  \sum^{N_2}_{i=1} \cfrac{x2_i}{\Delta y2_i^2}, \\
& S2_y = \sum^{N_2}_{i=1} \cfrac{y2_i}{\Delta y2_i^2},\\
&S2_{xx} = \sum^{N_2}_{i=1} \cfrac{x2^2_i}{\Delta y2_i^2},  &S2_{xy} =  \sum^{N_2}_{i=1} \cfrac{x2_i\ y2_i}{\Delta y2_i^2}. &
\end{align}

With this definitions Eq.~\eqref{eq:system} becomes:
\begin{equation}
\begin{cases}
a1\ S1 + b\ S1_x = S1_y \\
a2\ S2 + b\ S2_x = S2_y  \\ 
a1\ S1_x + b\ S1_{xx} + a2\ S2_x + b\ S2_{xx} = S1_{xy} + S2_{xy}
\end{cases}.
\end{equation}
The solution of these three equations with three unknowns is:
\begin{widetext}
\begin{equation}
\begin{cases}
a1 = -\cfrac{S2\ S1_x\ S1_{xy} + S2\ S1_x\ S2_{xy} + S2_x^2\ S1_y - S2\ S1_{xx}\ S1_y - S2\ S2_{xx}\ S1_y - S1_x\ S2_x\ S2_y}
{-S2\ S1_x^2 - S1\ S2_x^2 + S1\ S2\ S1_{xx}  + S1\ S2\ S2_{xx}}\\
a2 = -\cfrac{S1\ S2_x\ S1_{xy} - S1\ S2_x\ S2_{xy} + S1_x\ S2_x\ S1_y - S1_x^2\ S2_y + S1\ S1_{xx}\ S2_y + S1\ S2_{xx}\ S2_y}
{S2\ S1_x^2 + S1\ S2_x^2 - S1\ S2\ S1_{xx}  - S1\ S2\ S2_{xx}}\\ 
b = - \cfrac{S1\ S2\ S1_{xy} + S1\ S2\ S2_{xy} - S2\ S1_x\ S1_y - S1\ S2_x\ S2_y}{S2\ S1_x^2 + S1\ S2_x^2 - S1\ S2\ S1_{xx} - S1\ S2\ S2_{xx}}
\end{cases}.
\end{equation}
\end{widetext}



\section{Definition of the function $\boldsymbol{F(\Delta x, \Delta y, \Delta \Theta)}$} \label{app:F}
The exact form of $F$ in Eq.\ (\ref{defF}) can be deduced using simple algebraic equations and reference frame transformation 
formulas. 
If we take any pair of perpendicular lines in the reference CCD (see Fig.~\ref{ccd-rot}),
\begin{equation}
y = a_0 + b_0\ x   \quad \text{and} \quad x = c_0 + d_0\ y, \label{eq:lines-CCD-ref}
\end{equation}
the coordinates ($x_p$, $y_p$) from the line intersection  can be calculated  
on the selected CCD.
The parameters of these lines are deduced form the lines in the reference CCD
(Eq.~\eqref{eq:lines-CCD-ref}), taking into account the necessary change on 
$a_0$ and $c_0$ for the translation on the grating pattern:
\begin{equation}
\left( \begin{matrix}
x_p \\
y_p
\end{matrix} \right):
\begin{cases}
x_p = c_0 + sy + d_0\ y_p \\
y_p = a_0 + sx + b_0\ x_p 
\end{cases}.
\end{equation}
Here, $sx$ and $sy$ are the parameters of the translation that can be
easily deduced from the wafer image.
In this case we have:
\begin{equation}
\begin{cases}
x_p = \cfrac{c_0 + d_0 ( a_0 + sx)  + sy}{1- b_0\ d_0}\\
y_p = \cfrac{a_0 + b_0 (c_0 + sy) + sx}{1- b_0\ d_0}
\end{cases}.
\end{equation}

In the same way we can calculate the coordinates $(X_p,Y_p)$: the crossing point of the lines
in the selected CCD after the $\Delta \Theta$ rotation.
Before the rotation, the line coordinates on the selected CCD are:
\begin{equation}
y' = a + b\ x' \quad \text{and} \quad x' = c + d\ y'.
\end{equation}
After rotation around the CCD center $(X_C,Y_C)$ the line equations become (see Fig.~\ref{ccd-rot}):
\begin{equation}
Y = A + B X \quad \text{and} \quad X = C + D Y,
\end{equation} \small
where the line parameters are given by:
\begin{align} 
& A = \frac{a + b X_C - Y_C + (Y_C - b\ X_C) \cos \Delta \Theta - (X_C + b\ Y_C) \sin \Delta \Theta}{\cos \Delta \Theta - b \sin \Delta \Theta}\\
& B = \frac{b \cos \Delta \Theta + \sin \Delta \Theta} {\cos \Delta \Theta - b \sin \Delta \Theta}\\
& C = \frac{c - X_C + d\ Y_C + (X_C - d\ Y_C) \cos \Delta \Theta + (Y_C + d\ X_C) \sin \Delta \Theta}{\cos \Delta \Theta + d \sin \Delta \Theta}\\
& D = \frac{d \cos \Delta \Theta - \sin \Delta \Theta}{\cos \Delta \Theta + d \sin \Delta \Theta}.
\end{align}
With this reference change, the coordinates $(X_p,Y_p)$ are:
\begin{widetext}
\begin{equation}
\begin{cases}
X_p = \cfrac{(b\ d - 1)X_C - [c + a\ d + (b\ d - 1) X_C] \cos \Delta \Theta + (a + b\ c+ (b\ d - 1) Y_C) \sin \Delta \Theta}{b\ d -1}\\
Y_p = \cfrac{(b\ d - 1)Y_C - [c + a\ d + (b\ d - 1) Y_C] \cos \Delta \Theta + (a + b\ c+ (b\ d - 1) X_C) \sin \Delta \Theta}{b\ d -1}
\end{cases}.
\end{equation}
\end{widetext}

The function $F$ is defined as
\begin{multline}
F(\Delta x, \Delta y, \Delta \Theta) =  (X_p - x_p - \Delta x)^2 + (Y_p - y_p - \Delta y)^2\\
= \textbf{\Bigg(} \frac{1}{(b\ d - 1) (b_0\ d_0 - 1)} \\
\{ (b\ d - 1) (c_0 + a_0\ d_0 + d_0\ sx + sy -X_C + b_0\ d_0\ X_C) \\
- (b_0\ d_0 - 1)[c + a\ d + (b\ d -1) X_C] \cos \Delta \Theta \\
+ (b_0 d_0 - 1)[a + b\ c + (b\ d -1)Y_C] \sin \Delta \Theta \} - \Delta x \textbf{\Bigg)}^2 \\
+ \textbf{\Bigg(} \frac{1}{(b\ d - 1) (b_0\ d_0 - 1)} \\ 
\{ (b\ d - 1) (a_0 + b_0\ c_0 + b_0\ sy + sx - Y_C + b_0\ d_0\ Y_C) \\
- (b_0\ d_0 - 1)[a + b\ c + (b\ d - 1) Y_C] \cos \Delta \Theta \\
+ (b_0 d_0 - 1)[c + a\ d +  (b\ d - 1) Y_C]  \sin \Delta \Theta \} - \Delta y \textbf{\Bigg)}^2 .
\end{multline}




\newpage

\printtables

\newpage

\printfigures

\end{document}